\documentclass[aps,prl,twocolumn,showpacs,preprintnumbers,amsmath,amssymb]{revtex4-1}

\usepackage{graphicx}

\usepackage{dcolumn}
\usepackage{bm}
\begin{document}

\preprint{APS/123-QED}

\title{Electron cotunneling transport in gold nanocrystal arrays}

\author{H. Moreira$^1$, Q. Yu$^1$, B. Nadal$^1$, B. Bresson$^1$, M. Rosticher$^2$, N. Lequeux$^1$, A. Zimmers$^1$ and H. Aubin$^1$}
\email{Herve.Aubin@espci.fr}
\affiliation{(1) Laboratoire de Physique et d'Etude des Mat\'eriaux, UMR 8213, ESPCI-ParisTech-CNRS-UPMC, 10 rue Vauquelin, 75231 Paris, France\\
(2) Laboratoire Pierre Aigrain, CNRS, ENS, UPMC, 24 rue Lhomond, 75231 Paris, France\\
}

\date{\today}

\begin{abstract}
We describe current-voltage characteristics I(V) of alkyl-ligated gold nanocrystals $\sim 5~nm$ arrays in long screening length limit. Arrays with different alkyl ligand lengths have been prepared to tune the electronic tunnel coupling between the nanocrystals. For long ligands, electronic diffusion occurs through sequential tunneling and follows activated laws, as function of temperature  $\sigma \propto e^{-T_0/T}$  and as function of electric field $I \propto e^{-\mathcal{E}_0/\mathcal{E}}$. For better conducting arrays, i.e. with small ligands, the transport properties crossover to the cotunneling regime and follows Efros-Shklovskii laws as function of temperature $\sigma \propto e^{-(T_{ES}/T)^{1/2}}$ and as function of electric field $I \propto e^{-(\mathcal{E}_{ES}/\mathcal{E})^{1/2}}$. The data shows that electronic transport in nanocrystal arrays can be tuned from the sequential tunneling to the cotunneling regime by increasing the tunnel barrier transparency.
\end{abstract}

\maketitle

Arrays of metallic, semiconducting, or magnetic nanocrystals (NC) with radii of $2-7~nm$ can now be synthesized\cite{Talapin2010a}. Owing to their small self-capacitance, the charging energy for adding one electron per NC is large. Thus, these systems are ideally suited for the study of correlated electronic diffusion in presence of both disorder and strong Coulomb interactions\cite{Tran2005,Beloborodov2007}.

In weakly conducting arrays, the large density of states at the Fermi level of metallic NC limits hopping to nearest neighbors and the large Coulomb energy opens a hard gap for electronic transport. These characteristic features led to an activated temperature dependence of the conductance where electron transport occurs by sequential tunnel hopping\cite{Search2002}. While activated conductance was observed in granular metals\cite{Neugebauer1962} and in gold NC arrays at high temperature $T>100 K$ \cite{Parthasarathy2004,Wang2007}, the low temperature conductance of granular metals\cite{Sheng1973,Abeles1975,Gerber1997,Chui1981,Simon1987} and gold NC arrays\cite{Tran2005,Zabet-Khosousi2006,Tran2008,Sugawara2008, Nickels2008, Herrmann2007} has also been observed to follow the Efros-Shklovskii (ES) law :

\begin{equation}\label{hoppinglawtemp}
   \sigma_{ES} \simeq \sigma_0 exp[{-(T_{ES}/T)^{1/2}}]
\end{equation}


This variable range hopping law is usually observed in doped semiconductors\cite{Shklovskii1984} where the wavefunction of dopant states decays exponentially and long range tunneling between localized electrons is possible.

The observation of the ES law in granular metals is puzzling as it implies long range tunneling of electrons beyond nearest neighbors. This issue was recently addressed in a series of theoretical works \cite{Zhang2004,Feigelman2005,Beloborodov2005a,Beloborodov2007} which established that effective long range hopping is possible because of cotunneling, where the electron hops to large distance through a string of virtual charge states. Such cotunneling phenomena is well known from the study of the transport properties of individual quantum dots\cite{Aleiner2002}. In NC arrays, while the temperature dependence of the zero-bias conductance has been shown to follow Eq.~\ref{hoppinglawtemp} at low temperature, the question of the electric field dependence of electronic current in the cotunneling regime remains to be explored.

In this letter, we present measurements of I(V) characteristics of gold NC arrays in the limit of long screening length, where the inter-electrode distance is small and the capacitive coupling of the NC with the gate is small. In this limit, the laws describing electron transport in the regime of cotunneling are known, they follow ES type formula both for the temperature dependence, Eq.~\ref{hoppinglawtemp} and the electric field dependence :

\begin{equation}\label{hoppinglawfield}
   I_{ES} \simeq I_0 exp[{-(\mathcal{E}_{ES}/\mathcal{E})^{1/2}}]
\end{equation}

We study the evolution of the I(V) characteristics of the array as the electronic tunnel coupling between the NC is tuned by changing the ligand length. For small tunnel barrier transparency, the I(V) curves follow activated laws as function of temperature and electric field; for large tunnel barrier transparency, the I(V) curves follow ES laws, Eq.~\ref{hoppinglawtemp} at low temperature and  Eq.~\ref{hoppinglawfield} at low electric field.

\begin{figure}[h!]
\begin{center}
\includegraphics[width=8cm,keepaspectratio]{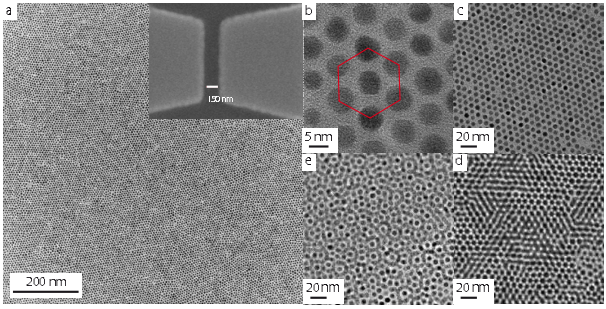}
\caption{ \label{Fig1} TEM image of dodecanethiol ligated gold NC self-organized in
compact hexagonal array deposited on TEM grids by the Langmuir-Schaefer method. Panel a) Gold NC monolayer shown on a large scale ($1~\mu m \times 1~\mu m$). Inset) Gold electrodes on which the array is deposited. Panel b) Zoom on the hexagonal compact array. Panels c, d, e) one monolayer, two monolayers and three monolayers, respectively.}
\end{center}
\end{figure}

Alkyl-ligated gold NC are synthesized by the digestive ripening method\cite{Lin2000}. This synthesis provides NC of radius $r\simeq 2.5~nm$ with dispersion less than $5\%$ in each batch. For this work, NC with alkylthiol-ligands of 6, 8 and 12 carbon atoms have been prepared, so-named $C6S$, $C8S$ and $C12S$ samples. The monodispersity of the NC allows for the preparation of hexagonal compact arrays with the Langmuir method\cite{Collier1997}. In a Langmuir trough filled with DI water, we first deposit a monolayer of dodecanethiol molecules. This amphiphile molecule improves the subsequent dispersion of gold NC. Then, a compact array is obtained upon compression of the barriers. This process is monitored optically and through measurements of the surface pressure.

After formation of the film, layers of NC are deposited with the Langmuir-Schaefer method; on substrates for conductivity measurements and on carbon grids for TEM characterization, shown Fig.~\ref{Fig1}. Two layers are usually deposited on the substrate to reduce pin-holes in the array. We did not find any significant differences in the transport properties as the number of layers is changed from two to four layers. The substrates are $p++$ doped silicon covered with a silicon oxide layer $500~nm$ thick. On these substrates, we fabricated $2~\mu m$ width gold electrodes separated of the length $L \simeq 150~nm$. See inset of Fig.~\ref{Fig1}. $C8S$ and $C6S$ samples were measured as deposited on the substrate. $C12S$ samples were dipped into isopropanol solution of alkyl-dithiol molecules during $5~min$ to cross-link the NC and increase the tunnel transparency. I(V) measurements as function of temperature were performed after cross-linking in solutions of octane-dithiol, hexane-dithiol, butane-dithiol and ethane-dithiol molecules, respectively called  $C8S2$, $C6S2$, $C4S2$ and $C2S2$ samples. Despite the quality of the array, disorder remains. First, electrostatic disorder due to random offset charges, second, tunnel barrier height disorder due to random ligands density and orientation. Both types of disorder are included within the theoretical model used below\cite{Beloborodov2005a} to analyze the data, where electrostatic disorder is modeled by grains with random potential and the tunnel matrix element is described by a random Gaussian variable.

\begin{figure}[h!]
\begin{center}
\includegraphics[width=8cm,keepaspectratio]{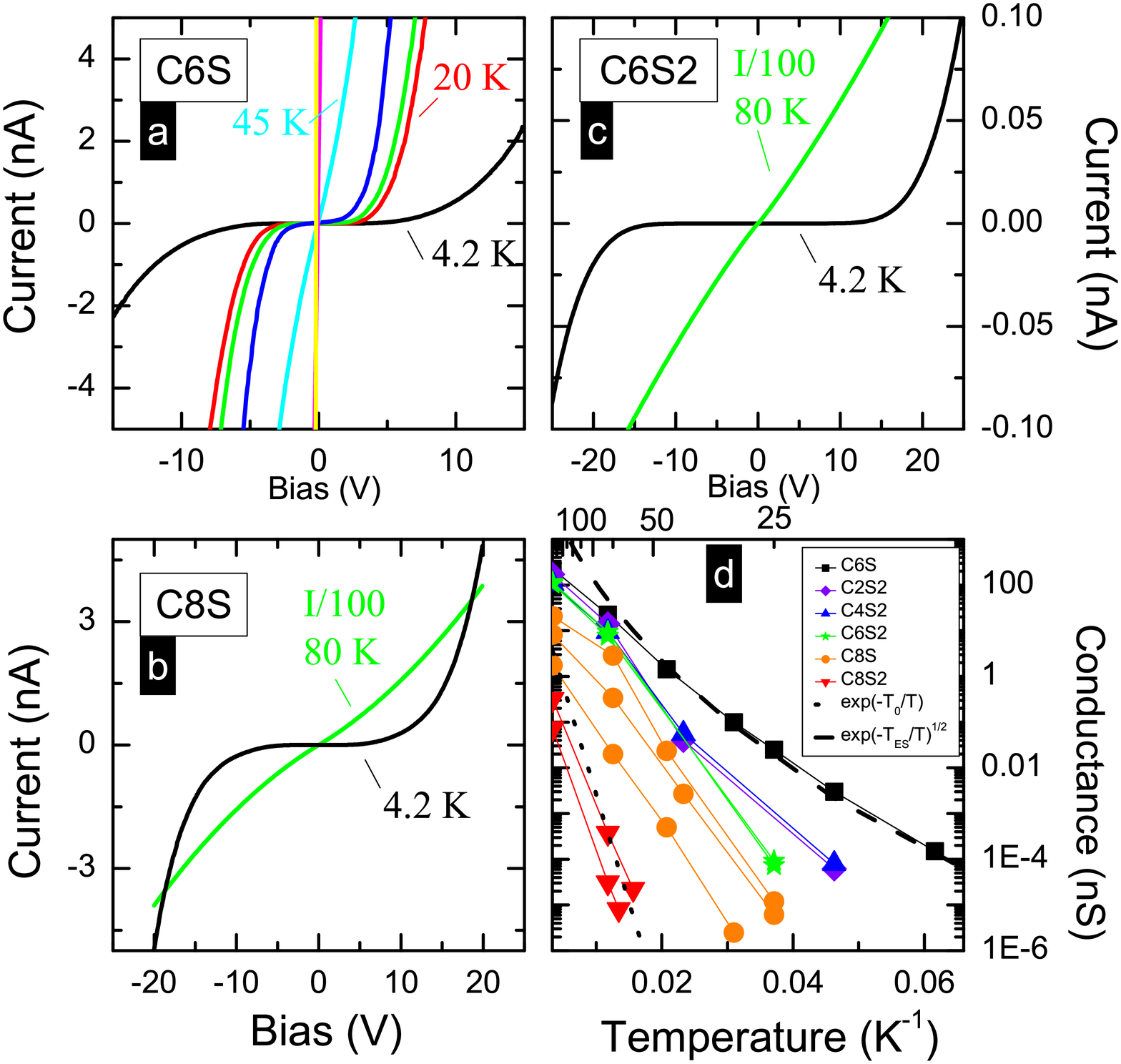}
\caption{ \label{Fig2} Panels a) \& b) I(V) curves of hexane-thiol (C6S) and octane-thiol (C8S) ligated NC arrays, respectively. Panel c) I(V) curve of dodecanethiol ligated NC crosslinked with hexanedithiol ligands (C6S2). Panel d) Zero-bias conductance of the arrays with varying ligand lengths. The weakly conducting arrays follow the activated law (dotted line). The more conducting arrays follow the ES law (dashed line). }
\end{center}
\end{figure}

Figure~\ref{Fig2} shows the I(V) curves of two as-deposited NC films $C8S$ and $C6S$ and one cross-linked film, $C6S2$. For all measured samples, the current increases non-linearly with voltage, as typically observed for NC arrays in the regime of Coulomb blockade\cite{Beloborodov2007}. Figure~\ref{Fig2}d shows the temperature dependence of the zero-bias conductance of all measured samples. Overall, the conductance of the arrays tends to increase as the ligands are shorter. This is expected as the conductance of N-alkanethiol ligands decreases exponentially fast with the number of carbon atoms $N$ as $G_N \simeq G_0 e^{-N}$\cite{Xu2003}, where $G_0=2e^2/h\simeq 77.5~\mu S$. The zero-bias conductance of weakly conducting samples as $C8S2$ and $C8S$ follows an activated temperature dependence $\sigma_{seq}\simeq \sigma_0 e^{-T_0/T}$ from room temperature to the lowest temperature ($T\simeq 20~K$) below which the conductance become smaller than the instrument sensitivity. The activation temperature $T_0$ is similar for all weakly conducting samples, it is set by the Coulomb energy $E_C=e^2/2C_0$ where $C_0=4 \pi \varepsilon \varepsilon_0 r \simeq 0.8~aF$ is the self-capacitance of the NC where the dielectric coefficient $\varepsilon=3$\cite{Tran2005}. Using those parameters, we find that the activation law with $T_0=E_C/k_B=1110~K$ can fit the zero-bias conductance of weakly conducting array as shown Fig.~\ref{Fig2}.

In contrast, the conductance of best conducting arrays with shorter ligands, $C4S2$, $C2S2$ and $C6S$, deviates from this activation law at low temperature, $T \lesssim 80~K$. This deviation cannot be due to a change in the activation energy. Theoretical calculations\cite{Beloborodov2005a} give for the activation energy of arrays of dimensionless conductance $g$, $E_A = E_C - (2gz/\pi)* E_{eh} \ln{2}$, valid when $gz<<1$, where $z=6$ is the coordination number and $E_{eh}=2E_C$ is the energy to create an electron-hole excitation in the system. As all the samples $CnS2$ are prepared from the same NC batch, there is no doubt that the NC size is the same for all arrays, as are the self-capacitance and the Coulomb energy $E_C$. Furthermore, as the conductivity of the arrays is very small, $g=G/G_0\simeq 10^{-4}$ at $T\simeq 80 K$ for the best conducting arrays, this provides only a very weak correction to the activation energy and consequently $E_A\simeq E_C$.

We show now that this deviation can be understood as a consequence of inelastic cotunneling transport of electrons\cite{Zhang2004,Feigelman2005,Beloborodov2005a}\footnote{Elastic cotunneling is also possible but at lower temperature and electric field range than measured here.}.
In this regime, the conductance should follow Eq.~\ref{hoppinglawtemp}. We find that this last formula fits the data with $T_{ES}=e^2/(4 \pi \varepsilon \varepsilon_0  <\xi> k_B) \simeq 8000~K$. The average value of the localization length $<\xi>=0.16 r$ is obtained from $\xi=\frac{2 r}{\ln{ [E_C^2/16 \pi (k_B T)^2 g ]}}$, valid when $g<<1$\cite{Beloborodov2005a}. This last formula shows that the localization length changes slowly with temperature and conductance : from $\xi\simeq 0.23 r$ ($T= 80~K$, $g=2\times10^{-4}$ ) to $\xi\simeq 0.09 r$ ($T = 5~K$, $g=4\times10^{-9}$ ).

We calculate that the hopping distance $r_{hop}=\sqrt{e^2 \xi/(8 k_B T\pi \varepsilon \varepsilon_0)}$, which increases with decreasing temperature, exceeds two NC below $T \sim 50~K$. This analysis confirms that electron transport occurs through cotunneling below that temperature and is in agreement with the conclusions of previous works by Tran et al. on similar gold NC arrays\cite{Tran2005}. We show now that an analysis of the electric field dependence of I(V) curves at low temperature allows to establish the signature of cotunneling transport in the array.



\begin{figure}[h!]
\begin{center}
\includegraphics[width=8cm,keepaspectratio]{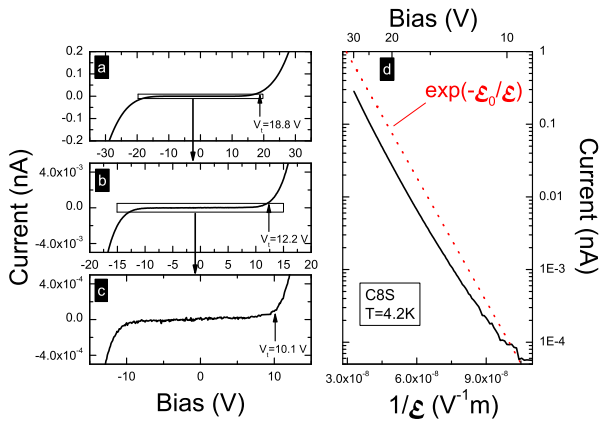}
\caption{ \label{Fig3} Panels a) to c) show that the I(V) curves are self-similar, i.e. the curves appear the same on different scales. This implies that the apparent voltage threshold $V_t$ indicated by arrows changes with the scale. This  shows that $V_t$ is not a relevant parameter for the transport properties of the array. In contrast, the Arrhenius plot -- panel d) -- shows that the electronic current follows an activated behavior $I(V)\propto e^{-\mathcal{E}_0/\mathcal{E}}$ (dotted line).}
\end{center}
\end{figure}

In previous works on gold NC arrays\cite{Tran2005, Tran2008, Sugawara2008, Nickels2008}, the non-equilibrium transport properties have been analyzed in the context of Middleton and Wingreen (MW) model\cite{Middleton1993} of percolating currents. One essential parameter of this model is the screening length $\lambda/r=C_0/C_G$, where $C_G=4\pi \varepsilon \varepsilon_0 r^2 /d$ is the capacitive coupling of the NC with the gate electrode. Short screening lengths, $\lambda/r \propto d/r\rightarrow 0$, are obtained in arrays strongly coupled to the gate electrode such as lithographically defined quantum dots \cite{Rimberg1995}. In this limit, the MW model predicts the existence of a voltage threesold $V_t$ for electron transport and a power law dependence of the current with voltage, $I\propto (V-V_t)^\alpha$. As the effects of cotunneling on this model are not known theoretically\cite{Beloborodov2007}, this prevents a reliable analysis of the effect of cotunneling on I(V) curves.

In our sample configuration, the screening length $\lambda\simeq d \simeq 500~nm$ is larger than the electrode separation $L=150~nm$, implying that the MW model does not apply\cite{Bascones2008}. In this long screening length regime, the laws describing the electric field dependence of the electronic current in the sequential and cotunneling regimes have been calculated theoretically\cite{Zhang2004,Beloborodov2005a}. They mirror the temperature dependence and can be obtained by replacing the thermal energy $\sim k_BT$ by the electrostatic energy $\sim eV$.

Experimentally, while an activated electric field dependence has been found for the conductance of granular metals\cite{Sheng1972}, there is no report of the observation of the ES law Eq.~\ref{hoppinglawfield} for the electric field dependence.

Figure \ref{Fig3} shows data for a $C8S$ sample. On left panels, the data is shown on a linear plot at three different scales of increasing magnification where arrows indicate for every I(V) curve the apparent voltage threshold $V_t$. It clearly appears that $V_t$ decreases upon increasing the plot magnification, which shows that $V_t$ is not a relevant physical parameter. On the right panel, the same data on an Arrhenius plot shows that the current follows the activated law for sequential tunneling, $I_{seq} \simeq I_0 e^{-\mathcal{E}_0/\mathcal{E}}$ where $\mathcal{E}_0=k_BT_0/e l$ is set by the Coulomb energy, $l$ is the inter-nanocrystal distance on which the electron is submitted to the electric field, $\mathcal{E}=cV/L$, where $c$ takes into account the screening effects that reduce the actual electrical field from the nominal value $V/L$. A good fit of the data can be obtained with $c=0.15$ and $l=1~nm$, giving for the activation electric field $\mathcal{E}_0=1.3\times10^8~Vm^{-1}$. While there is a large uncertainty on the actual electric field applied on the electron because of the uncertainty on inter-nanocrystal distance $l$ and on screening effects, we will see now that using the same values for the parameters $\mathcal{E}_0$ and $c$, the ES law can fit the data in the cotunneling regime.



\begin{figure}[h!]
\begin{center}
\includegraphics[width=8cm,keepaspectratio]{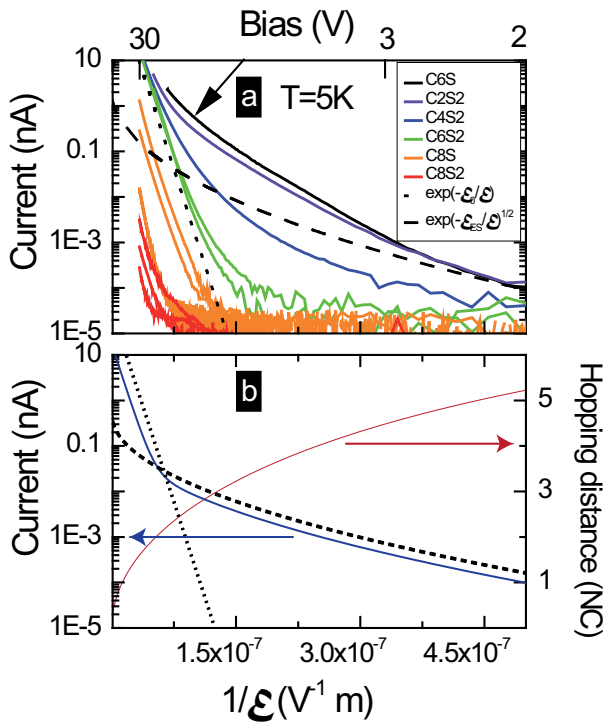}
\caption{ \label{Fig4} Panels a) shows the evolution of the I(V) curves of the arrays as the tunnel transparency between NC increases. For weakly conducting NC arrays, i.e. small tunnel barrier transparency, the I(V) curve follows an activated law, dotted line. As the conductance of the arrays increase, i.e. for larger tunnel barrier transparency, the I(V) curve follows an ES-type law, dashed line. The arrow indicates the field scale where the crossover between the two regimes occurs. Panel b) The thin red line is the hopping distance. The dotted line is the activated law, the dashed line is the ES law. As the hopping distance increases at low electric field, a crossover occurs from the activated law to ES law, as shown schematically by the thin blue line.}
\end{center}
\end{figure}

Figure~\ref{Fig4} shows the I(V) curves for all samples. The I(V) curves of weakly conducting samples, $C8S2$ and $C8S$, follow the activated electric field dependence just described on the whole measured voltage range. In contrast, the I(V) curves of best conducting arrays, $C4S2$, $C2S2$ and $C6S$, deviate from this law at low electric field. It follows the ES law Eq.\ref{hoppinglawfield} with the same characteristic electric field $\mathcal{E}_{ES}=\mathcal{E}_{0}=1.3\times10^8~Vm^{-1}$. Upon increasing the voltage, a crossover is observed at an electric field $\mathcal{E} \simeq 10^7~Vm^{-1}$ and the I(V) curve recovers the activation law at high electric field. This is a remarkable observation. We find that the electric field dependence mirrors the temperature dependence and that a single characteristic electric field value, set by the Coulomb energy, can describe the I(V) curves for a set of NC arrays whose room temperature conductance changes by three orders of magnitude.

The same figure shows that the calculated hopping distance, $r_{hop}=\sqrt{e /(\mathcal{E} 4 \pi \varepsilon \varepsilon_0)}$\cite{Beloborodov2005a} increases rapidly about the electric field value where the crossover between the activated law and the ES law is observed. This confirms our analysis, the crossover occurs between the cotunneling regime at low electric field, where the hopping distance $r_{hop} \simeq 5 \times r$ at zero-bias, and the sequential tunneling regime, where $r_{hop} \simeq r$ at high bias.

To summarize, we studied the I(V) curves of gold NC arrays with long screening length. In this regime, the effects of cotunneling on the I(V) curves are known theoretically, whose predictions can be compared to the data. As a function of temperature, the conductance follows an activated law at high temperature in the sequential tunneling regime and an ES law at low temperature in the cotunneling regime. We find that the electric field dependence mirrors this temperature dependence. The I(V) curves follow an activated field dependence at high electric field in the sequential tunneling regime and an ES law at low electric field in the cotunneling regime. The crossover from the activated law to the ES law as array conductance is increased can be understood from the early calculations for linear NC chains where it was shown\cite{Averin1990,Geerligs1990} that the cotunneling current across $j$ junctions increases with the junction conductance $G$ as $I\propto [\frac{G}{G_0}]^j$.

We acknowledge useful discussions with B. Dubertret, M. Feigel'man, A. Ioselevich, T. Pons and V. Vinokur.
We acknowledge support from ANR programs "QUANTICON" and "CAMELEON".

\bibliography{goldBiblio}

\end{document}